\documentclass[pdflatex,sn-mathphys-num]{sn-jnl}

\usepackage{graphicx}%
\usepackage{multirow}%
\usepackage{amsmath,amssymb,amsfonts}%
\usepackage{amsthm}%
\usepackage{mathrsfs}%
\usepackage[title]{appendix}%
\usepackage{xcolor}%
\usepackage{textcomp}%
\usepackage{manyfoot}%
\usepackage{booktabs}%
\usepackage{algorithm}%
\usepackage{algorithmicx}%
\usepackage{algpseudocode}%
\usepackage{listings}%
\usepackage{booktabs}
\usepackage{makecell}
\usepackage[version=4]{mhchem}
\usepackage[numbers]{natbib}

\theoremstyle{thmstyleone}%
\theoremstyle{thmstyletwo}%

\theoremstyle{thmstylethree}%

\begin{document}

\title[Article Title]{A Hierarchical Synergistic Deep Learning Framework Integrating Composition, Structure, and Ionic Transport for Solid-State Electrolyte Discovery}


\author[1,2,4]{\fnm{Hongwei} \sur{Du}}

\author[1,2]{\fnm{Dingyang} \sur{Lv}}

\author[1,2]{\fnm{Baole} \sur{Wei}}

\author[1,2]{\fnm{Yongheng} \sur{Li}}

\author[1,2]{\fnm{Feng} \sur{Yu}}

\author*[1,2,3]{\fnm{Ziheng} \sur{Lu}}

\author*[5]{\fnm{Siqi} \sur{Shi}}

\author*[4]{\fnm{Hong} \sur{Wang}}
\email{hongwang2@sjtu.edu.cn}

\affil[1]{\orgname{Zhongguancun Academy},\orgaddress{\city{ Beijing},\postcode{ 100094},\country{ China}}}

\affil[2]{\orgname{Zhongguancun Institute of Artificial Intelligence},\orgaddress{\city{ Beijing},\postcode{ 100094},\country{ China}}}

\affil[3]{\orgname{Kairos Materials},\orgaddress{\city{ Beijing},\postcode{ 100094},\country{ China}}}

\affil[4]{\orgdiv{School of Materials Science and Engineering},\orgname{ Shanghai Jiao Tong University},\orgaddress{\city{ Shanghai},\postcode{ 200240},\country{ China}}}

\affil[5]{\orgdiv{ State Key Laboratory of Materials for Advanced Nuclear Energy \& School of Materials Science and Engineering},\orgname{ Shanghai University},\orgaddress{\city{ Shanghai},\postcode{ 200444},\country{ China}}}


\abstract{Inorganic solid-state electrolytes require simultaneous satisfaction of multiple performance constraints including high room-temperature ionic conductivity, wide electrochemical window, excellent electronic insulation, and favorable mechanical compliance. Single modeling approaches struggle to support reliable multi-objective screening within ultra-large chemical spaces due to inherent challenges including training data distribution mismatch, cross-property dataset heterogeneity, and kinetic transport data scarcity. To address these limitations, a hierarchical synergistic deep learning screening framework is developed to address these limitations through sequential coordination of computational efficiency, prediction accuracy, and screening reliability. This framework integrates four functionally complementary modules in a tiered manner. The in-house-developed L-G-DCNN and a multi-fidelity implementation built on DenseGNN serve as compositional and structural experts for thermodynamic coarse screening and multi-property evaluation, respectively, while MatterSim and system-specific DeePMD models are further employed for transport pre-assessment and kinetic validation. Systematic benchmark studies demonstrate that each module achieves superior performance relative to mainstream counterparts in its respective task domain, and retrospective validation establishes a dual closed-loop verification system covering both module-level model accuracy and full-workflow reliability. The framework is deployed for large-scale screening of 30,364,908 Alex/ICSD-derived candidates, ultimately identifying 97 high-performance candidates with room-temperature ionic conductivities ranging from 0.109 to 59.0 mS/cm, including 94 halides, one borohydride, and two oxides. Consistency analysis with independent experimental data confirms that 76 out of the 94 halide candidates fall within previously reported high-conductivity structural regions. Further analysis reveals that Li\textsuperscript{+} jump network connectivity, rather than the quantity of geometric Li sites, acts as the core factor determining room-temperature ionic conductivity. Li defect engineering can effectively enhance the transport performance of oxides, whereas the inherent structural rigidity of the O\textsuperscript{2-} framework suggests a potential upper limit on oxide electrolyte performance.}

\maketitle

\section{Introduction}\label{sec1}

Inorganic solid-state electrolytes (SSEs) require simultaneous possession of high room-temperature ionic conductivity, a broad electrochemical window, superior electronic insulation and stable mechanical compliance~\cite{1,2,3}. Over the past decade, computational discovery of SSEs has evolved from density-functional-theory (DFT) database mining and bond-valence or topology-based transport descriptors to machine-learning-assisted high-throughput screening and machine-learning-potential (MLP)-enabled molecular dynamics (MD)~\cite{4,5,6}. Materials Project (MP) and ICSD-derived workflows have enabled increasingly systematic evaluations of phase stability, electrochemical stability and electronic insulation, while interpretable machine learning and graph neural networks (GNNs) have accelerated the prioritization of Li-containing compounds across known structural databases~\cite{5,7,8}. More recently, prototype expansion, generative crystal design and universal or domain-oriented MLP have further shifted SSE discovery from manual chemistry-by-analogy toward data-driven exploration of much larger hypothetical spaces and from purely static descriptors toward dynamic Li-ion transport validation~\cite{9,10,11,12,13,14,15}. However, expanding candidate generation from known databases to large hypothetical spaces also shifts the bottleneck from finding possible structures to reliably ranking them under coupled thermodynamic, electrochemical, mechanical and finite-temperature transport constraints~\cite{4,10,14}.
\begin{figure}[t]
    \centering
    \includegraphics[width=0.7\columnwidth]{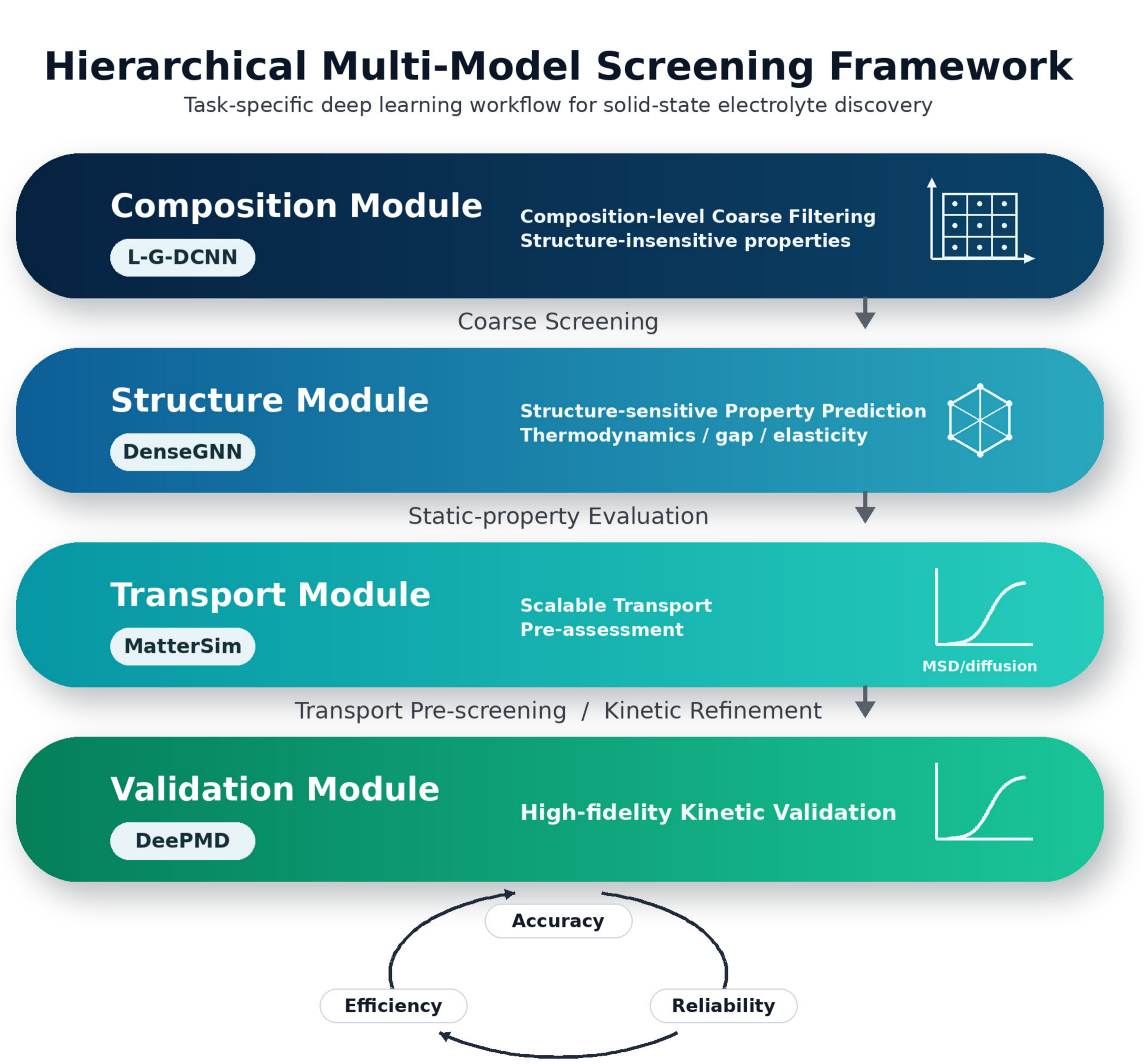}
    \caption{Hierarchical synergistic screening framework for materials discovery. The workflow integrates specialized models across sequential screening stages to balance computational efficiency, predictive accuracy, and validation reliability. L-G-DCNN serves as the front-end compositional module for cost-effective coarse filtering of ultra-large candidate spaces. DenseGNN, the structural module, evaluates structure-sensitive static properties via multi-fidelity data fusion. MatterSim assesses Li\textsuperscript{+} mobility through MD-derived diffusion behavior, and DeePMD provides high-precision kinetic validation for top candidates. This staged architecture narrows the candidate pool while progressively increasing physical fidelity from compositional screening to dynamic transport validation.}
    \label{fig:figure1}
\end{figure}

Despite these methodological advances, the competing demands of computational efficiency, prediction accuracy, and screening reliability expose inherent hurdles that complicate multi-constraint SSE screening.  First, prototype substitution strategies can rapidly generate hundreds of millions of unrelaxed raw structures, many of which are thermodynamically unstable. Most training data for existing structure-based GNNs come from stable or metastable structures, leading to generalization issues~\cite{12,16}. Second, reliable SSE screening requires simultaneous prediction of multiple key properties including thermodynamic stability, electronic insulation, electrochemical window and mechanical moduli. However, the sizes of training datasets for different properties can differ by up to three orders of magnitude, and their chemical coverage is uneven~\cite{17}. Such dataset heterogeneity complicates joint model training and motivates the multi-fidelity strategy adopted here. Third, kinetic labels remain scarce and strongly time dependent, while static structure-to-property GNNs~\cite{18,19,20,21,22} primarily learn migration descriptors rather than directly resolving finite-temperature ionic conductivity. Recent benchmarks further reveal substantial, chemistry-dependent errors in migration-barrier and trajectory predictions~\cite{14,23,24}.

To address these limitations, a hierarchical synergistic screening workflow is designed that strategically distributes distinct prediction tasks across specialized models operating within their individually validated performance ranges. This cascaded architecture assigns task-specific models with distinct representations to successive screening stages, balancing computational cost, property-prediction accuracy, and transport-validation reliability. The first module adopts the L-G-DCNN compositional model, a previously developed compositional deep learning model~\cite{25}, for thermodynamic coarse screening of tens of millions of initial candidates. As the front-end screening step, compositional feature-based rapid elimination of thermodynamically unstable structures at minimal computational cost avoids accuracy degradation of structure-dependent models on out-of-distribution configurations. The second module employs a multi-fidelity implementation built on the previously developed DenseGNN graph neural network~\cite{18} to realize multi-objective evaluation of structure-sensitive properties including convex hull energy, band gap and elastic moduli. Fusion of training data with varying computational accuracy effectively alleviates inherent dataset heterogeneity. The third module implements MatterSim for structural relaxation, phase diagram analysis and room-temperature ionic conductivity assessment~\cite{26}. Explicit kinetic simulation compensates for the intrinsic defects of static property prediction models. The fourth module conducts long-term kinetic refinement on limited candidate structures via DeePMD to guarantee the credibility of final screening outputs. Each tier of the framework is tailored to address a specific bottleneck discussed above. The L-G-DCNN compositional module mitigates generalization degradation on unrelaxed structures. The multi-fidelity DenseGNN implementation alleviates dataset heterogeneity through data fusion. The MatterSim-DeePMD cascade addresses the kinetic prediction limitations inherent to static machine learning models (Fig.~\ref{fig:figure1}).

To systematically validate the efficacy of the proposed framework, hierarchical model configuration is performed tailored to the inherent characteristics of compositional, structural, and dynamic data domains. Each hierarchical module is rigorously benchmarked to ensure reliable prediction accuracy and generalization performance. Retrospective validation of classic SSEs is further conducted under relaxed screening criteria, and the multi-dimensional computational outputs are well consistent with established experimental trends. These outcomes collectively establish a dual closed-loop validation system that verifies both module-level model accuracy and full-workflow reliability. Building on this, is deployed the framework to screen 30,364,908 unlabeled Alex/ICSD-derived candidates, ultimately obtaining 97 high-performance candidate structures with room-temperature ionic conductivities exceeding 0.1 mS/cm. Systematic analysis reveals the regulatory effects of anion types on material properties, as well as the elemental composition, crystal system distribution, and structural motif-conductivity correlation characteristics of the final candidate set, establishing the decisive chemical advantage of halides under multi-property constraints. Subsequently, the 94 halide candidates were projected onto the cation polarization descriptor space from independent experimental literature~\cite{27} to verify the rationality of the screening results and quantify the overlap between calculated and experimental values. Further topological analysis of the Li\textsuperscript{+} jump networks of representative oxide, halide and borohydride materials reveals a core transport-governing principle that room-temperature ionic conductivity relies on the connectivity of Li\textsuperscript{+} jump networks rather than the quantity of geometric Li sites. With the low-transport oxide Li\textsubscript{3}BO\textsubscript{3} serving as the model system, the effectiveness and trade-offs of four modification strategies including Li defect engineering, B-site doping, anion regulation and high-entropy configuration are systematically investigated. Li defect engineering can effectively boost the transport performance of oxides, while the inherent structural rigidity of oxide anion frameworks imposes a potential upper limit on their ionic transport performance.
\section{Results}\label{sec2}

\subsection{Systematic Validation and Analysis of the Screening Framework}

\begin{figure}[t]
    \centering
    \includegraphics[width=\columnwidth]{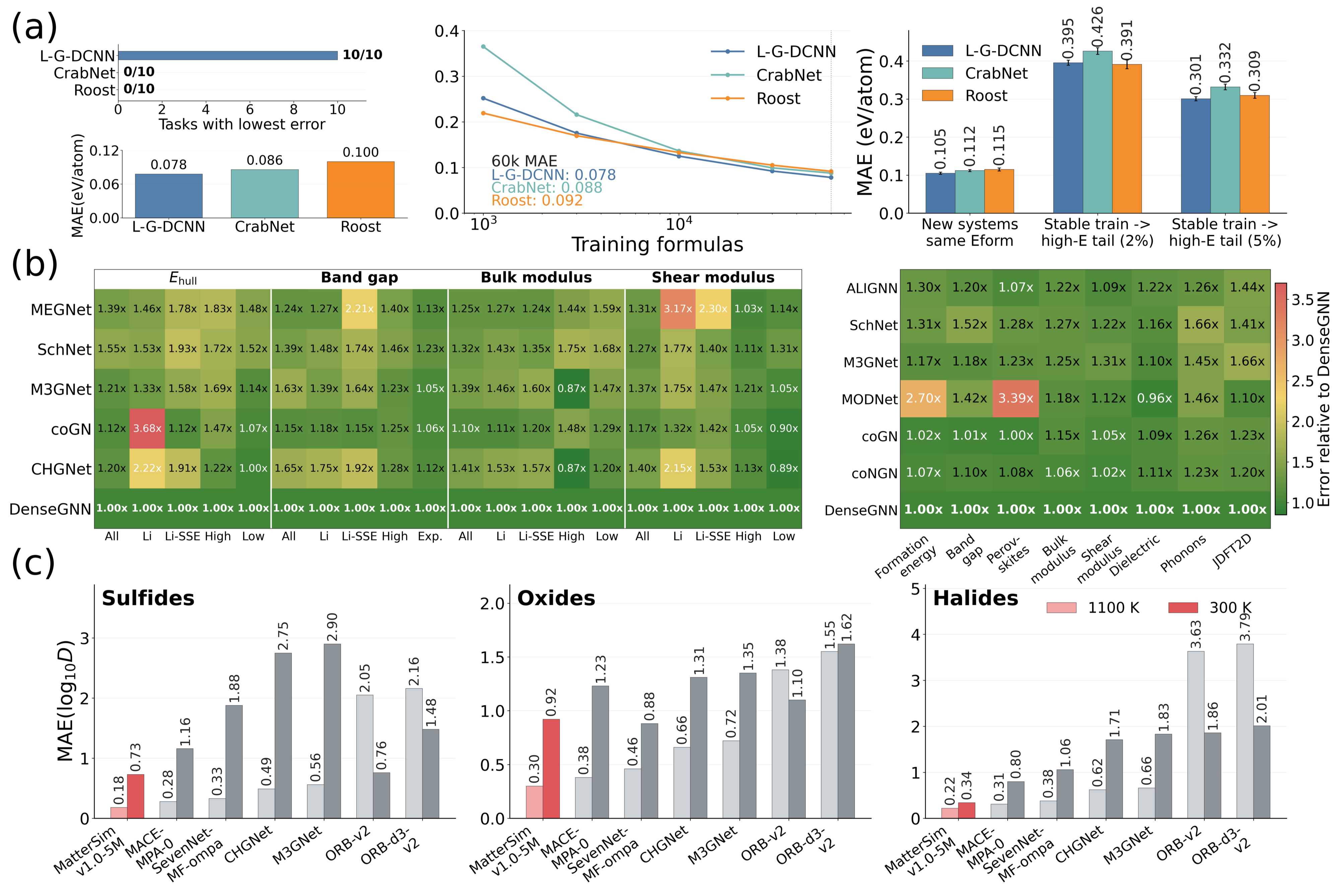}
    \caption{Quantitative benchmarks of the models used in the hierarchical screening workflow. (a) Composition-based L-G-DCNN benchmarks. The left panel reports the number of Matbench composition-regression tasks for which each model achieves the lowest mean absolute error (MAE). The middle panel shows formation-energy MAE as a function of the number of training formulas on a logarithmic x-axis. The right panel shows five-fold cross-validated MAEs for three composition-only extrapolation tests, with error bars denoting standard deviations. (b) Structure-based DenseGNN benchmarks. The left heatmap summarizes errors for convex-hull energy, band gap, bulk modulus, and shear modulus across multi-scale validation, tail-extrapolation, and experimental-generalization tests. The right heatmap shows normalized Matbench MAE ratios relative to DenseGNN, with DenseGNN set to 1.00. (c) Machine-learning-potential benchmarks for Li-ion diffusion. Bars show mean absolute logarithmic errors of Li\textsuperscript{+} diffusion coefficients relative to DeePMD references for sulfide, oxide, and halide systems at 1100 K and 300 K.}
    \label{fig:figure2}
\end{figure}

Each tier of the hierarchical framework is first validated through systematic benchmarks before deployment to the Alex/ICSD extended space. Fig.~\ref{fig:figure2} addresses three core research questions raised in the introduction and systematically validates the rationality of model selection within the proposed screening framework across three tiers including compositional coarse screening, structural multi-property prediction, and ionic transport evaluation, which delivers quantitative evidence for defining the performance limits of the entire screening workflow. The three subplots separately characterize the generalization performance of the compositional screening model, the multi-property prediction accuracy of graph neural networks, and the reliability of kinetic potentials for transport evaluation. Rather than simply stacking a single universal model, this hierarchical architecture assigns dedicated models tailored to the task-specific features of compositional, structural and dynamic data domains and mitigates the inherent performance drawbacks of generic models under multi-constraint screening. Fig.~\ref{fig:figure2}\textcolor{blue}{(a)} verifies the suitability of L-G-DCNN for preliminary compositional screening~\cite{25}. Benchmarked against composition-agnostic regression tasks from the Matbench dataset~\cite{28}, L-G-DCNN achieves the lowest MAE across all evaluation tasks. It yields an MAE of 0.078 eV/atom for formation enthalpy prediction on MP data, outperforming the Roost and CrabNet models~\cite{29,30}. Learning efficiency tests using formula-only inputs reveal marginal superiority of Roost under small training datasets. As the training set expands to 60 000 chemically diverse formulas, L-G-DCNN attains a reduced test MAE of 0.078 eV/atom and surpasses the other two comparative models. Two formula-based distribution extrapolation experiments further confirm the competitive performance of L-G-DCNN in cross-chemical-system generalization and stable-unstable extrapolation with a 5\% high-energy anchor fraction. Roost only exhibits slightly better prediction accuracy under the extreme tail condition with a low anchor ratio of 2\%. Balancing prediction precision and computational overhead, L-G-DCNN meets the requirements of early-stage thermodynamic coarse screening over candidate pools containing tens of millions of materials.

Fig.~\ref{fig:figure2}\textcolor{blue}{(b)} validates the capability of DenseGNN for predicting structurally sensitive multi-properties~\cite{18}. The left heatmap compares the predictive performance of six mainstream structural models on four key material properties namely convex hull energy, band gap, bulk modulus and shear modulus. The test setups cover sampled MP datasets, lithium-bearing compound subsets, lithium solid-state electrolyte subsets and extrapolation scenarios. Additional ICSD experimental datasets are incorporated into band gap evaluation to examine generalization toward experimental measurements. All heatmap values are normalized relative to the MAE of DenseGNN. The right Matbench benchmark~\cite{28} encompasses eight representative material property prediction tasks. Most mainstream architectures including ALIGNN, SchNet and M3GNet~\cite{31,32,33} exhibit larger prediction errors than DenseGNN, while only MODNet~\cite{34} delivers comparable accuracy for dielectric property calculation. Accordingly, DenseGNN is deployed to conduct multi-objective screening of structurally sensitive properties such as convex hull energy, band gap and elastic moduli.

Before presenting the MLP benchmark results in Fig.~\ref{fig:figure2}\textcolor{blue}{(c)}, it is important to clarify the role of DeePMD as the system-specific reference used in this study~\cite{35,36}. The DeePMD framework constructs a many-body neural-network potential energy surface trained on first-principles energies, forces, and virials, while preserving the fundamental symmetries of atomistic systems. Extensive work has established that DeePMD can reproduce quantum-mechanical reference data with linear-scaling molecular dynamics efficiency. When training configurations are systematically generated through concurrent-learning protocols such as DP-GEN~\cite{37}, the fidelity of a DeePMD potential should be established against held-out DFT energies, forces, and stresses rather than assumed a priori. For the system-specific models used here, the relevant energy–force–stress validation is reported in Fig. S10–S12. Fig.~\ref{fig:figure2}\textcolor{blue}{(c)} presents benchmark results for identifying optimal MLP dedicated to transport calculations. Three archetypal SSEs families including sulfides, oxides and halides are examined. Long-time DeePMD molecular dynamics simulations with supercells exceeding 1200 atoms serve as the study-specific benchmark, and mean absolute logarithmic diffusion error is adopted as the evaluation metric. Among the evaluated models, MatterSim~\cite{26} produces the minimal prediction error for all three material families at 1100 K. At 300 K, MatterSim maintains optimal accuracy for sulfide and halide systems and achieves comparable performance to SevenNet-MF-ompa~\cite{38} for oxides, outperforming widely used potentials such as MACE-MPA-0, ORB, CHGNet and M3GNet overall~\cite{14,33,39,40,41}. While newer MLPs with an order-of-magnitude larger parameter count may offer slightly higher predictive accuracy, computational efficiency is explicitly factored into model selection to meet the high-throughput demands of the hierarchical synergistic framework. Guided by these benchmark results, MatterSim is adopted as the cost-optimal choice for structural relaxation, phase diagram calculation and room-temperature transport pre-evaluation throughout the screening workflow.

\begin{figure}[t]
    \centering
    \includegraphics[width=0.6\columnwidth]{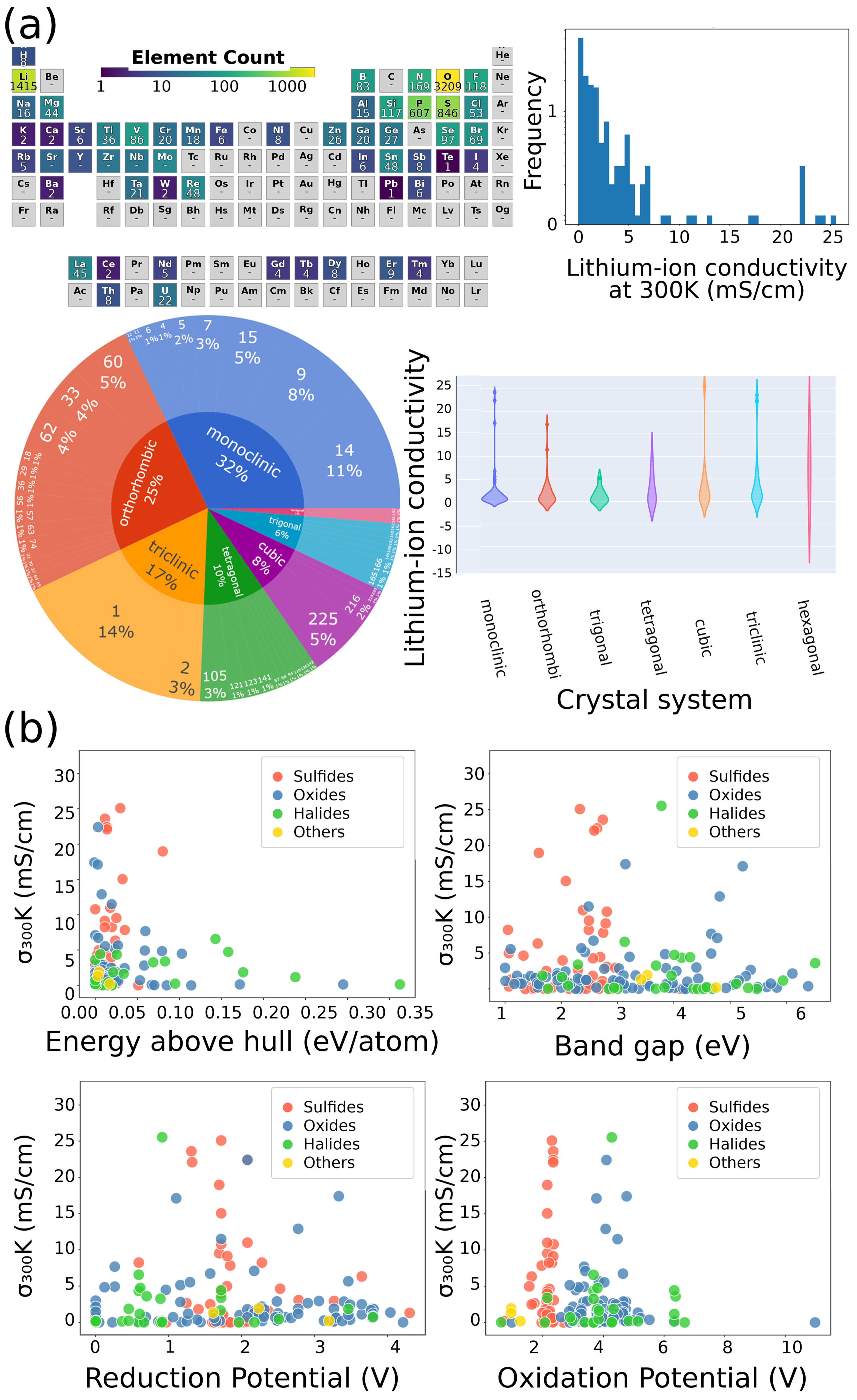}
    \caption{Analysis of Common SSE Families in MP with Relaxed Constraints. (a) Statistics of elemental abundance, 300 K Li\textsuperscript{+} conductivity distribution, space group/crystal system distribution, and violin plots of conductivity grouped by crystal system for 156 SSEs. The horizontal axes represent elements, conductivity ranges, and space groups/crystal systems, respectively, and the vertical axes correspond to counts, frequencies, or Li\textsuperscript{+} conductivities. (b) Correlations between 300 K Li\textsuperscript{+} conductivity and convex hull energy, band gap, reduction potential, and oxidation potential. The horizontal axes represent the corresponding properties, the vertical axis represents Li\textsuperscript{+} conductivity, and colors distinguish sulfides, oxides, halides, and other materials. This figure is used to verify the consistency between the screening workflow and known chemical trends.}
    \label{fig:figure3}
\end{figure}
Prior to large-scale screening in the Alex/ICSD extended space, a rigorous same-protocol benchmark system was constructed on 170,470 DFT-labeled compounds from the MP database~\cite{5} to quantitatively validate the decision accuracy and reliability of the workflow. Systematic comparison with a single-model M3GNet control workflow~\cite{33} demonstrates substantially higher decision agreement with DFT reference results at critical screening nodes, particularly for ionic conductivity where false-positive rates are markedly reduced. Detailed screening funnel comparisons, decision agreement statistics, and MAE reduction data across properties are provided in Supporting Information (SI) Fig. S1-S7.
\begin{table*}[t]
    \centering
    \caption{Four candidate materials retained by the present strategy that
    are consistent with the DFT reference under the same Materials Project
    (MP) benchmark protocol.}
    \label{tab:candidate_materials}

    \resizebox{\linewidth}{!}{%
        \begin{tabular}{lccccccc}
            \toprule
            Formula
            & \makecell{Space\\group}
            & \makecell{Energy above hull\\(eV atom$^{-1}$)}
            & \makecell{Band gap\\(eV)}
            & \makecell{Voltage window\\(V)}
            & \makecell{Bulk modulus\\(GPa)}
            & \makecell{Shear modulus\\(GPa)}
            & \makecell{Conductivity\\(mS cm$^{-1}$)}
            \\
            \midrule

            \ce{Li4SiO4}
            & P1
            & 0.0303
            & 5.2439
            & 3.098
            & 77.437
            & 46.417
            & 0.931
            \\

            \ce{LiK2NbO4}
            & P1
            & 0.0371
            & 3.8068
            & 2.726
            & 33.055
            & 19.423
            & 0.891
            \\

            \ce{Li10Mg7Cl24}
            & Cm
            & 0.0108
            & 4.5584
            & 3.387
            & 29.427
            & 13.716
            & 0.125
            \\

            \ce{Li3YBr6}
            & C2/c
            & 0.0256
            & 4.1363
            & 3.083
            & 16.264
            & 12.767
            & 4.847
            \\

            \bottomrule
        \end{tabular}%
    }
\end{table*}

The small number of candidates retained under strict criteria reflects the scarcity of known materials in the MP database that simultaneously satisfy all multi-property constraints. To further validate the generalizability of the screening framework across common SSE chemistries and examine whether established chemical trends are correctly reproduced, screening thresholds were relaxed to include representative sulfide, halide, and oxide families. Under these relaxed criteria, common SSE families such as sulfides, halides, and oxides in the MP database were screened, ultimately yielding 156 materials. Fig.~\ref{fig:figure3}\textcolor{blue}{(a)} shows the elemental composition distribution, indicating that the materials are mainly composed of alkali metals (e.g., Li), alkaline earth metals, some transition metals, post-transition metals, and non-metals (O, N, F, Cl, Br, Si, P, S, Se, Ge), consistent with the composition of SSEs reported in the literature~\cite{1,2}. The ionic conductivity distribution shows that most conductivities are below 5.0 mS/cm, with a few ranging from 10 to 25 mS/cm, matching the literature range. The crystal-system statistics show that the monoclinic system accounts for the largest share (\~32\%), followed by the orthorhombic (\~25\%) and triclinic (\~17\%) systems, while the cubic, tetragonal, trigonal, and hexagonal systems account for approximately 8\%, 6\%, 5\%, and 3\%, respectively (percentages do not sum exactly to 100\% owing to rounding). This distribution reflects the dominance of low-symmetry monoclinic and orthorhombic frameworks among common SSEs. The violin plots show the distribution characteristics of Li\textsuperscript{+} conductivity at 300 K for different crystal systems, revealing the potential influence of crystal system symmetry on ion migration ability and providing structural design directions for screening high-conductivity SSEs. Fig.~\ref{fig:figure3}\textcolor{blue}{(b)} presents the relationships between ionic conductivity and convex hull energy, band gap, reduction potential, and oxidation potential for the 156 candidate materials, categorized by halides, oxides, and sulfides. The vast majority of materials have convex hull energies below 0.1 eV/atom, indicating high thermodynamic stability, with only a few oxides and halides as exceptions. The band gaps of the sulfides are mainly concentrated below 2.5 eV, with few exceeding 3.0 eV, consistent with the intrinsically narrow band gaps of sulfide chemistries. In contrast, the band gaps of the oxides and halides are more widely distributed and show no obvious trend, reflecting their greater chemical diversity. Materials with reduction potentials below 1.0 V exhibit good compatibility with existing anodes. Among them, halides generally have reduction potentials less than 1.0 V, showing better reductive stability, while sulfides have reduction potentials concentrated in the range of 1.0 to 2.5 V, resulting in poor compatibility that requires further optimization. Oxides show no obvious trend in reduction potential. The oxidation potentials of oxides and halides are generally higher than 3.0 V, showing good compatibility with existing cathode materials, while the oxidation potentials of sulfides are mostly below 2.5 V, making them difficult to match the voltage range of high-voltage cathodes~\cite{2,27,42,43,44}. These results align with the characteristics of common SSEs, confirming the reliability of the screening workflow.

\subsection{Extended Alex/ICSD Application Workflow and Candidate Category Analysis}
\begin{figure}[t]
    \centering
    \includegraphics[width=\columnwidth]{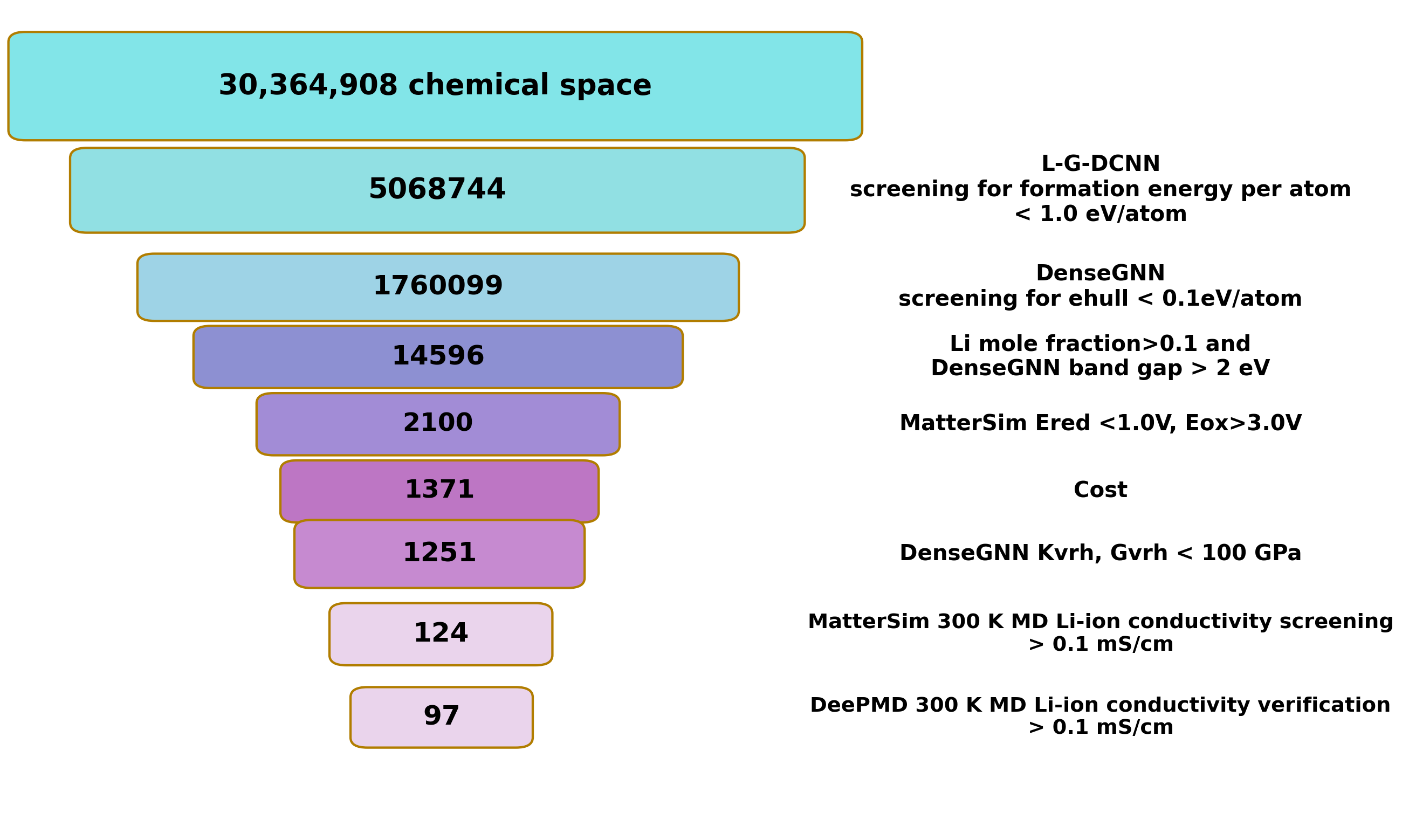}
    \caption{Multi-stage screening workflow for the Alex/ICSD-derived candidate space. The funnel reports the number of retained structures after each sequential filter, decreasing from 30,364,908 initial candidates to 97 DeePMD-validated room-temperature SSE candidates. The thresholds were arranged from low-cost permissive stability screening to increasingly stringent thermodynamic, electronic, electrochemical, mechanical, and kinetic criteria, so that high-fidelity transport validation was applied only to the most promising late-stage candidates.}
    \label{fig:figure4}
\end{figure}
The hierarchical screening framework was applied to a large Alex/ICSD-derived chemical space constructed from the Alexandria computational library and 2022 ICSD experimental structures~\cite{7,12} through elemental equivalent substitution, comprising 30,364,908 candidate structures. As summarized in Fig.~\ref{fig:figure4}, the workflow follows a cost-aware sequence in which inexpensive coarse filters first remove chemically implausible candidates, while structure-sensitive and kinetic evaluations are reserved for progressively smaller candidate pools. In the first stage, L-G-DCNN was used for composition-level coarse screening based on periodic-table descriptors and stoichiometric features. A permissive formation-energy threshold of $<$ 1.0 eV/atom was adopted to reject clearly unstable chemistries without prematurely discarding potentially metastable SSE candidates, reducing the pool to 5,068,744 structures. The second stage introduced DenseGNN-based structure-sensitive screening. A convex-hull-energy threshold of $<$ 0.1 eV/atom imposed a metastability constraint~\cite{45}, followed by Li mole fraction $>$ 0.1 and band gap $>$ 2.0 eV to select Li-containing electronic insulators, yielding 14,596 candidates. The third stage evaluated battery applicability through electrochemical, elemental, and mechanical constraints. MatterSim/Pymatgen electrochemical-window analysis~\cite{26,42,43,46} retained candidates satisfying reduction $<$ 1.0 V and oxidation $>$ 3.0 V. Element availability/cost filtering and DenseGNN elastic-modulus criteria, KVRH/GVRH $<$ 100 GPa, further removed impractical or mechanically stiff structures, leaving 1,251 candidates. Finally, because static models cannot directly resolve Li-ion transport, kinetic screening was performed using MD-based validation. MatterSim 300 K MD identified 124 candidates with Li-ion conductivity $>$ 0.1 mS/cm, and DeePMD long-time 300 K MD with trajectory quality control confirmed 97 final high-conductivity room-temperature SSE candidates.
\begin{figure}[t]
    \centering
    \includegraphics[width=\columnwidth]{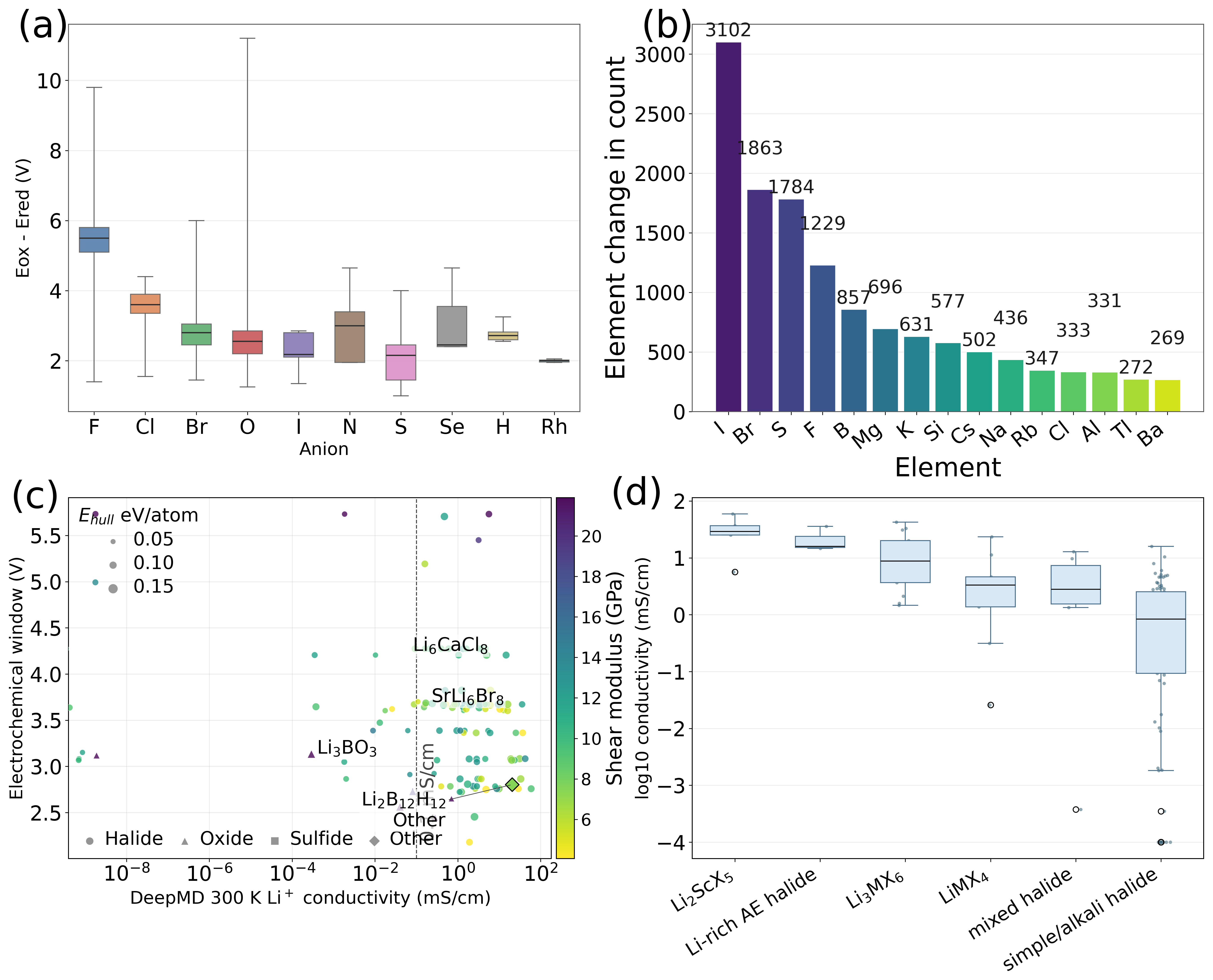}
    \caption{Chemical Trends and Candidate Structure Property from Alex/ICSD Screening. (a) 0 K electrochemical window distribution grouped by anions for 2,100 candidates passing the band gap and electrochemical window thresholds. The horizontal axis represents anion elements, and the vertical axis represents window width, comparing the effects of different anions. (b) Changes in elemental counts as the voltage window threshold increases from 1.0 V to 1.5 V and 2.0 V. The horizontal axis represents elements, and the vertical axis represents occurrence counts. (c) Four-dimensional property map of the 124 candidate structures. The horizontal axis represents DeePMD 300 K Li\textsuperscript{+} conductivity, the vertical axis represents electrochemical stability window, the dashed line indicates the 0.1 mS/cm threshold, and point color/size corresponds to convex hull energy and shear modulus, respectively. (d) Correlation between halide structural motifs and conductivity. The horizontal axis represents structural motif categories, and the vertical axis represents log\textsubscript{10}(conductivity).}
    \label{fig:figure5}
\end{figure}

Fig.~\ref{fig:figure5} integrates the anion dependence of electrochemical windows, the comprehensive property map of the 124 screened structures, and the transport differences of halide structural motifs, providing quantitative statistical analysis to systematically reveal the chemical enrichment patterns and property distribution characteristics of the candidate structures. Fig.~\ref{fig:figure5}\textcolor{blue}{(a)} compares the 0 K electrochemical windows of different anion systems based on 2,100 qualified candidates: halides exhibit the optimal window widths, followed by oxides, while chalcogenides generally have narrower windows. This result is consistent with experimental observations: halides have better electrochemical window and can match high-voltage cathodes~\cite{2,27,44}. Sulfides have outstanding ionic transport capabilities but insufficient oxidation potentials. Fig.~\ref{fig:figure5}\textcolor{blue}{(b)} reflects the changes in elemental composition as the electrochemical window threshold increases: halogen elements are significantly enriched in the high-window region, and boron also maintains a high proportion throughout, indicating that boron-based structures are potential candidates for developing SSEs with wide electrochemical windows. Fig.~\ref{fig:figure5}\textcolor{blue}{(c)} presents the multi-property map of the 124 MatterSim pre-screened structures after DeePMD re-evaluation. False-positive samples were strictly eliminated false-positive samples based on MSD curve characteristics and goodness of fit, ultimately obtaining 97 high-performance candidates with room-temperature conductivities $>$ 0.1 mS/cm, comprising 94 halides, 2 oxides, and 1 closo-type borohydride~\cite{47} (a metastable cubic polymorph of Li\textsubscript{2}B\textsubscript{12}H\textsubscript{12}). Their conductivities range from 0.109 to 59.0 mS/cm. Fig.~\ref{fig:figure5}\textcolor{blue}{(d)} shows that, within the present candidate set, flexible frameworks containing large-radius, polarizable halogens are associated with more continuous Li\textsuperscript{+} migration channels, consistent with reported halide structure–transport trends. Most notably, a prominent enrichment of high-performance candidates within halide chemistries is observed across all screening tiers. This key finding is further validated through direct comparison with independent experimental literature data.
\subsection{Experimental Literature Consistency Verification }
\begin{figure}[t]
    \centering
    \includegraphics[width=\columnwidth]{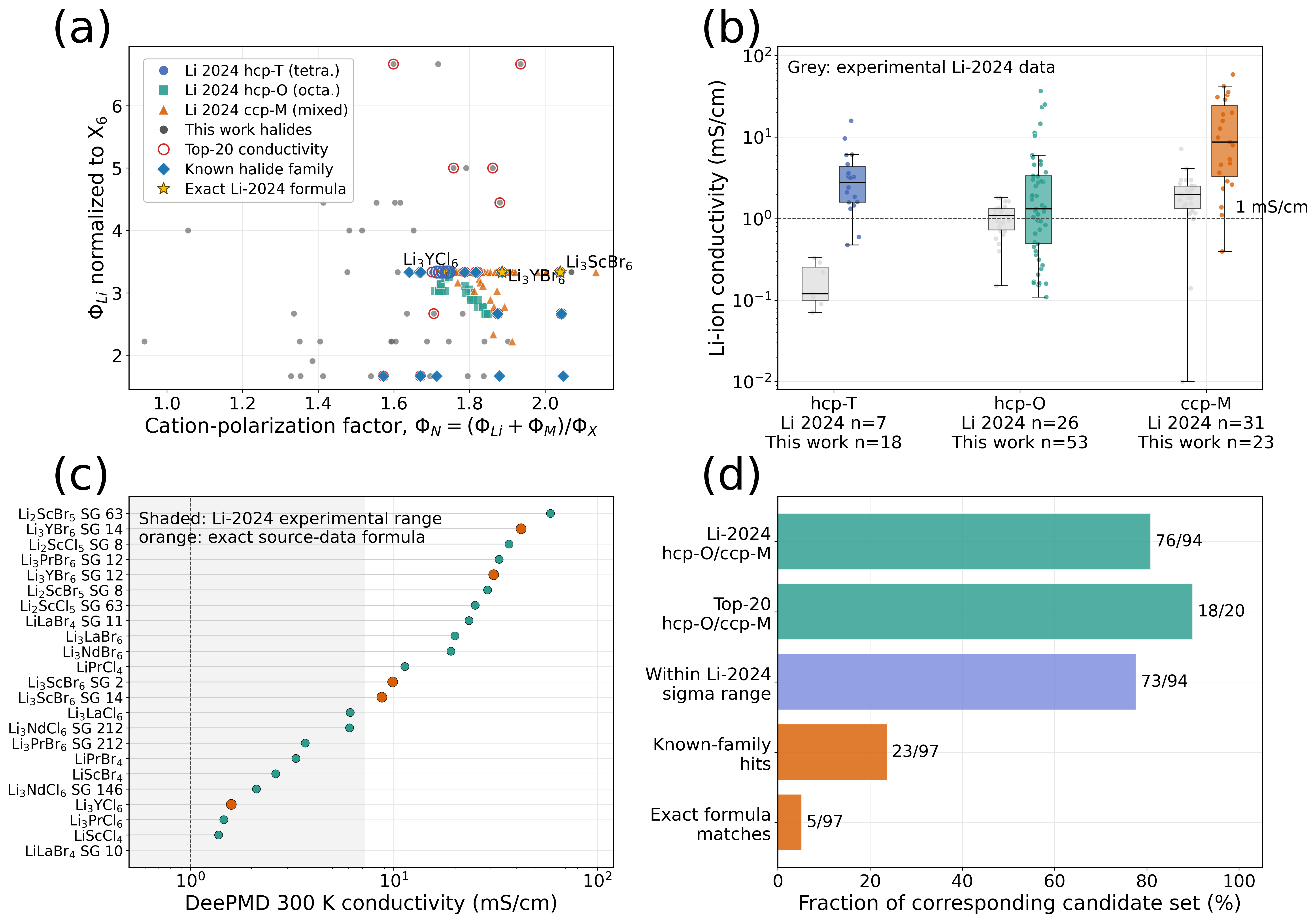}
    \caption{Quantitative Comparison of the 97 Final Candidates with Halide Experimental Data. (a) Comparison in the cation polarization descriptor space. The horizontal axis represents $\phi_N = \frac{\phi_{\mathrm{Li}^{+}} + \phi_M}{\phi_X}$, and the vertical axis represents \textsubscript{6}-coordination normalized ${\phi_{Li}}$. In the Li 2024 dataset, hcp/ccp denote approximate hexagonal close-packed and cubic close-packed halogen anion sublattices, respectively, while T/O/M represent tetrahedral, octahedral, and mixed cation occupancy. Blue circles, green squares, and orange triangles correspond to literature points for hcp-T, hcp-O, and ccp-M phases, respectively. Gray points represent halide candidates from this work, red open circles denote the top 20 candidates by conductivity, blue diamonds indicate candidates overlapping with known halide families, and gold stars mark candidates with identical chemical formulas to those in the Li 2024 literature dataset. (b) Conductivity range comparison with Li-2024 data in nearest-neighbor literature regions. The horizontal axis represents nearest-neighbor literature structure regions, the vertical axis represents Li\textsuperscript{+} conductivity, gray represents Li 2024 experimental data, and colored represents DeePMD-calculated values. (c) Conductivity distribution of candidates overlapping with known families. The horizontal axis represents DeePMD 300 K conductivity, the vertical axis represents candidate chemical formulas and space group, and the gray shaded area represents the experimental dataset range. (d) Statistical hierarchy of experimental consistency evidence. The horizontal axis represents the fraction of the corresponding candidate set, and the vertical axis represents evidence categories.}
    \label{fig:figure6}
\end{figure}

Screening results reveal prominent enrichment of high-performance candidates among halides. To assess whether this enrichment is chemically reasonable rather than an artifact of the screening models, an independent 2024 experimental halide SSE dataset reported by Li et al. was used as an external consistency reference for the 94 halide candidates. Oxide and borohydride candidates were excluded from this projection to avoid extending halide-specific empirical descriptors to other structural frameworks. Projection analysis is conducted using the cation polarization descriptor system proposed by Li et al (2024)~\cite{27} with the formula $\phi_N = \frac{\phi_{\mathrm{Li}^{+}} + \phi_M}{\phi_X}$ and X\textsubscript{6}-coordination normalized ${\phi_{Li}}$. The descriptor enables clear differentiation of three phase regions: hcp-T (hexagonal close-packed with tetrahedral cation occupancy), hcp-O (hexagonal close-packed with octahedral cation occupancy), and ccp-M (cubic close-packed with mixed cation occupancy). The latter two regions match experimentally confirmed domains with high ionic conductivity. In Fig.~\ref{fig:figure6}\textcolor{blue}{(a)}, the results show that 76 out of 94 halide candidates fall within the experimentally verified hcp-O or ccp-M high-conductivity phase regions, including 18 of the top 20 candidates by conductivity. This indicates that the high-performance halides screened in this work are consistent with known high-conductivity materials. Fig.~\ref{fig:figure6}\textcolor{blue}{(b)} further compares the conductivity ranges within the nearest-neighbor experimental structure regions, showing that the calculated conductivities of 73 out of 94 candidates fall within the measured ranges of the corresponding experimental regions. Fig.~\ref{fig:figure6}\textcolor{blue}{(c)} presents the conductivity distribution of final candidates associated with reported halide SSE families. Among the 97 final candidates, 23 show clear structural or compositional relationships to reported halide SSE families, and 5 have identical chemical formulas to experimentally reported materials. For the remaining candidates unassigned to known families or exact-formula matches, formula-level cross-referencing against experimental literature and accessible materials databases found no direct experimental precedent for most examined halide compositions. These candidates are therefore categorized as formula-level unreported halide SSE predictions from our workflow, while definitive crystallographic novelty requires further prototype- and polymorph-level matching against curated crystallographic databases. Fig.~\ref{fig:figure6}\textcolor{blue}{(d)} summarizes the statistical hierarchy of literature-consistency evidence, including descriptor-region agreement, conductivity-range overlap, known-family association, and exact-formula matches. These comparisons support the consistency of the screened halide candidates with experimentally reported halide SSE trends, but they do not constitute direct experimental validation of the predicted structures. For candidates whose computed conductivities exceed the upper bound of experimentally reported values in the corresponding chemical family, including the Li\textsubscript{2}B\textsubscript{12}H\textsubscript{12} polymorph and the highest-conductivity bromides, these values should be treated as priority targets for re-examining the synthesizability of the specific polymorph, and the influence of phase purity, rather than as confirmed ultra-high-conductivity materials.

An independent static transport-descriptor cross-check was performed using CAVD+BVSE~\cite{48,49,50} (Fig. S13). Across the 123 BVSE-applicable structures in the broad validation pool, the three-dimensional BVSE percolation barrier was inversely associated with the 300 K conductivity ranking. The association remained significant within the final cohort, while the overlap between high-conductivity and low-barrier rankings provided a complementary top-k consistency measure. All 96 BVSE-applicable final candidates were completed. Li\textsubscript{2}B\textsubscript{12}H\textsubscript{12} was outside the adopted ionic BVSE applicability domain because its covalent B–H cluster prevents an unambiguous oxidation-state assignment. BVSE is therefore interpreted as an independent static ranking-consistency test, not as a substitute for finite-temperature MD or AIMD.

\subsection{Representative Material Structural Analysis and Oxide Design}

\begin{figure}[t]
    \centering
    \includegraphics[width=\columnwidth]{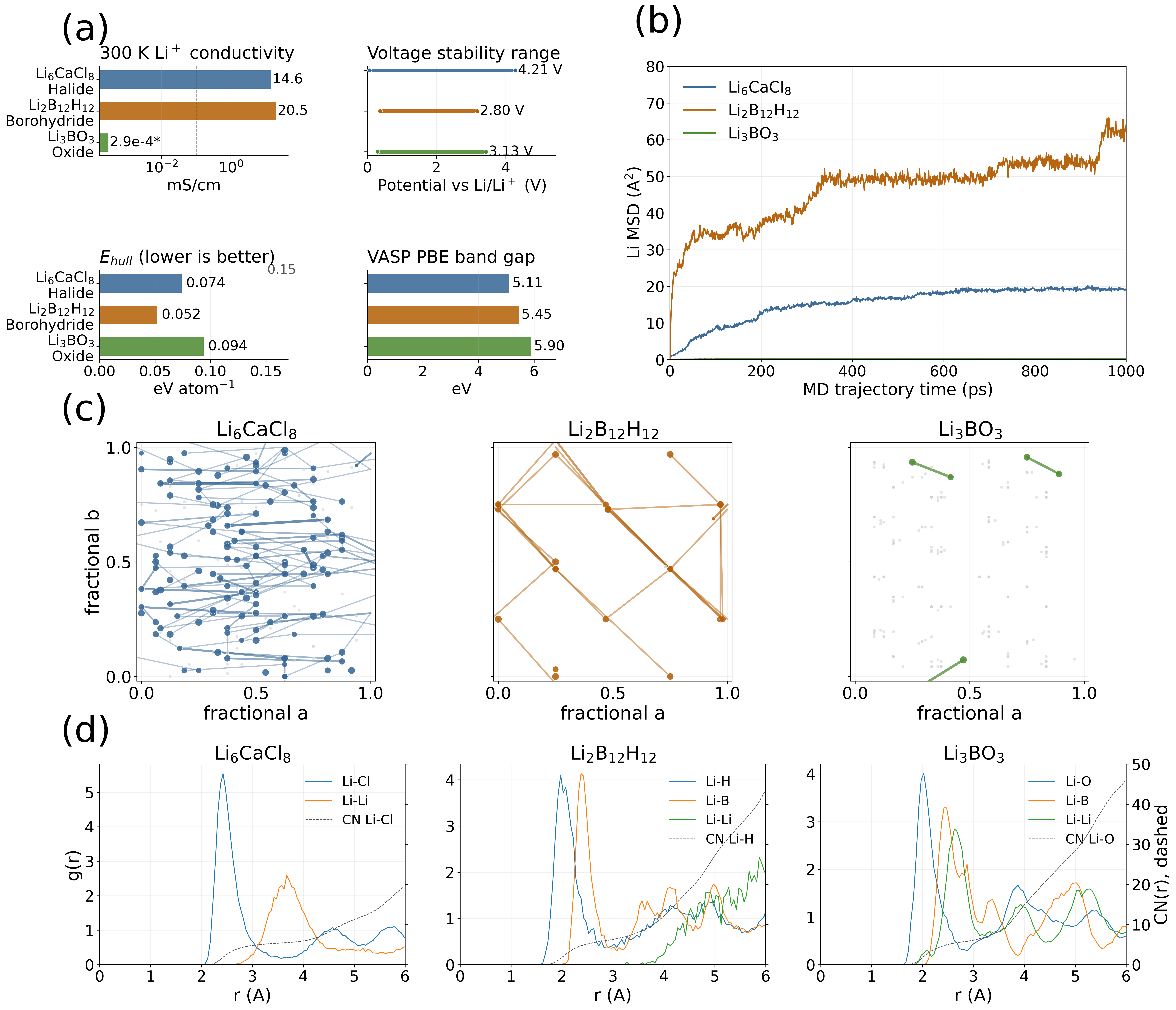}
    \caption{Comparison of DeePMD Transport Mechanisms for the Three Representative Materials. (a) Comparison of 300 K Li\textsuperscript{+} conductivity, electrochemical window, Ehull, and VASP PBE band gap for Li\textsubscript{6}CaCl\textsubscript{8}, Li\textsubscript{2}B\textsubscript{12}H\textsubscript{12}, and Li\textsubscript{3}BO\textsubscript{3}. (b) 0–1000 ps Li\textsuperscript{+} MSD curves. The horizontal axis represents MD time, the vertical axis represents Li MSD. (c) Jump-active Li site networks. Coordinates are fractional a and b. Gray represents non-jumping sites, and colored nodes and connections represent the actual jump networks from the 1 ns trajectory. (d) Partial radial distribution function (partial RDF)/CN(r) centered on Li. The horizontal axis represents distance r, the left vertical axis represents g(r), and the right vertical axis (dashed line) represents CN(r).}
    \label{fig:figure7}
\end{figure}
To elucidate the microscopic nature of Li\textsuperscript{+} transport in different anion frameworks and provide a theoretical baseline for modifying low-transport oxides, three representative electrolytes were selected for comparison. These include the halide Li\textsubscript{6}CaCl\textsubscript{8}, the borohydride Li\textsubscript{2}B\textsubscript{12}H\textsubscript{12}, and the oxide Li\textsubscript{3}BO\textsubscript{3}. It should be emphasized that closo-hydroborates are an experimentally well-established solid-electrolyte family rather than a new chemistry identified in this work~\cite{47}. The Li\textsubscript{2}B\textsubscript{12}H\textsubscript{12} candidate examined here is a metastable cubic polymorph (space group Fm-3, No. 202), which is structurally distinct from the experimentally characterized ordered $\alpha$-phase (Pa-3). Experimentally, ordered $\alpha$-Li\textsubscript{2}B\textsubscript{12}H\textsubscript{12} undergoes an order–disorder transition near 615 K. Reported room-temperature conductivities are strongly preparation-dependent: pristine or dried $\alpha$-phase samples are typically in the 10\textsuperscript{-8}–10\textsuperscript{-7} S/cm range, a mechanically milled sample has been reported at 3.1 × 10\textsuperscript{-4} S/cm, and the equimolar Li\textsubscript{2}B\textsubscript{10}H\textsubscript{10}–Li\textsubscript{2}B\textsubscript{12}H\textsubscript{12} mixed-anion phase reaches 4 × 10\textsubscript{-4} S/cm yet the effect of ball milling is not uniform across reports~\cite{51,52,53}. The simulation-derived value of 20.5 mS/cm therefore applies only to this specific modeled metastable polymorph within the present potential and trajectory protocol. Selection focused on the typical structures and transport characteristics of each system, with Li\textsubscript{3}BO\textsubscript{3} serving as a unified reference for subsequent defect and doping modification studies. Fig.~\ref{fig:figure7}\textcolor{blue}{(a)} reveals performance trade-offs among the three materials. Li\textsubscript{2}B\textsubscript{12}H\textsubscript{12} exhibits the highest room-temperature conductivity of 20.5 mS/cm but a narrow electrochemical window of only 2.80 V. Li\textsubscript{6}CaCl\textsubscript{8} shows the best balanced performance, with a Nernst-Einstein tracer/self-diffusion estimated conductivity on the same order of magnitude as benchmark halides (14.6 mS/cm, comparable to experimental Li\textsubscript{3}GdCl\textsubscript{3}Br\textsubscript{3} (11 mS/cm) and Li\textsubscript{3}YBr\textsubscript{3}Cl\textsubscript{3} (7.2 mS/cm)~\cite{54,55}), supported by trajectory quality metrics (self-MSD R\textsuperscript{2}=0.987, charge-MSD R\textsuperscript{2}=0.943), paired with an ultra-wide window of 4.21 V. Li\textsubscript{3}BO\textsubscript{3} possesses a wide band gap of 5.90 eV and a stability window of 3.13 V, but its conductivity is only 2.9 × 10\textsuperscript{-3} mS/cm, nearly five orders of magnitude lower. All three materials have convex hull energies below 0.1 eV/atom. Fig.~\ref{fig:figure7}\textcolor{blue}{(b)} shows the evolution of Li\textsuperscript{+} mean squared displacement (MSD) over a 1 ns simulation at 300 K for the three materials. The Li\textsuperscript{+} MSD of Li\textsubscript{2}B\textsubscript{12}H\textsubscript{12} and Li\textsubscript{6}CaCl\textsubscript{8} increases show sustained growth over the 1 ns trajectories, indicating active Li\textsuperscript{+} migration and long-range transport behavior within the sampled time window. The final 1 ns MSD value of Li\textsubscript{2}B\textsubscript{12}H\textsubscript{12} reaches 64.0 \AA\textsuperscript{2}, approximately 3.3 times that of Li\textsubscript{6}CaCl\textsubscript{8} (19.1 \AA\textsuperscript{2}), consistent with its higher conductivity. In contrast, the MSD of Li\textsubscript{3}BO\textsubscript{3} plateaus early in the simulation, with a final 1 ns value of only 0.18 \AA\textsuperscript{2}, confirming negligible effective long-range Li\textsuperscript{+} migration at room temperature. Fig.~\ref{fig:figure7}\textcolor{blue}{(c)} visualizes Li site networks where jumps occurred during the 1 ns trajectory, revealing that jump network connectivity, rather than the total number of geometrically identifiable Li sites, is the core structural factor governing room-temperature conductivity. Statistics indicate that of the 210 Li ions in Li\textsubscript{6}CaCl\textsubscript{8}, 105 underwent at least one jump, and 134 sites contributed to the jump network, enabling continuous migration channels across the supercell. Although Li\textsubscript{2}B\textsubscript{12}H\textsubscript{12} contains only 40 Li sites, 18 underwent jumps and 23 sites contributed to the network, yielding excellent channel connectivity. In contrast, of the 240 Li sites in Li\textsubscript{3}BO\textsubscript{3}, only 3 underwent a single jump and 5 sites were activated, failing to form any long-range connected migration paths. This directly explains the poor room-temperature transport performance of Li\textsubscript{3}BO\textsubscript{3} despite its high Li content. Fig.~\ref{fig:figure7}\textcolor{blue}{(d)} presents the partial radial distribution functions (RDFs) and coordination numbers (CNs) centered on Li. For Li\textsubscript{6}CaCl\textsubscript{8}, the first Li-Cl peak is located at approximately 2.43 \AA, reflecting the dynamic flexibility of the halide coordination environment. The Li-Cl CN saturates rapidly at short distances, and Li-Li correlations only accumulate significantly beyond 3.5–4.0 \AA, corresponding to the abundant Li site jump edges in the 3–4 \AA range in Fig.~\ref{fig:figure7}\textcolor{blue}{(c)}. This connectable medium-range structure under the flexible Cl framework enables stable long-range Li\textsuperscript{+} diffusion, consistent with the continuous linear MSD growth and conductivity of 14.6 mS/cm. For Li\textsubscript{2}B\textsubscript{12}H\textsubscript{12}, the Li-H/Li-B nearest-neighbor coordination is complete. The key distinction lies in the medium-range coordination environment. Li-Li correlations remain weak below 4 \AA but develop a broad feature at 5-6 \AA, indicating that the soft closo-[B\textsubscript{12}H\textsubscript{12}]\textsuperscript{2-} framework supports long-distance Li jumps through a flexible medium-range coordination environment. This interpretation is consistent with the median jump distance of 5.04 \AA in Fig.~\ref{fig:figure7}\textcolor{blue}{(c)}, the steep MSD growth, the final MSD value of 64.0 \AA\textsuperscript{2}, and the room-temperature conductivity of 20.5 mS/cm. For Li\textsubscript{3}BO\textsubscript{3}, the first Li-O peak is sharp (2.03 \AA), and the nearest-neighbor coordination number is not low. However, the CNs for Li-O, Li-B, and Li-Li all increase rapidly and continuously in the 3–6 \AA range. Although Li-Li sites are geometrically dense, they are strongly localized by the compact Li-B-O medium-range shell. This corresponds to only 3 Li sites undergoing single jumps and the failure to form a connected jump network in Fig.~\ref{fig:figure7}\textcolor{blue}{(c)}, resulting in negligible long-range Li\textsuperscript{+} migration. This is consistent with the rapid MSD plateauing, the low final MSD value of 0.18 \AA \textsuperscript{2}, and the extremely low room-temperature conductivity of 2.9 × 10\textsuperscript{-4} mS/cm.

To extend the representative mechanism analysis to the complete screened cohort, trajectory-level diagnostics were compiled for all 97 final candidates (Fig. S8 and S9). The same 300 K, 1 ns trajectories and a common post-processing protocol were used to calculate PBC-unwrapped Li MSD curves and density-site jump networks. Displaying every trajectory exposes the heterogeneous transport signatures underlying the scalar conductivity ranking, including approximately linear diffusion, intermittent jumps, and plateau-like behavior. The accompanying networks report the number of active Li ions, jump events, connected-component size, and spanning directions, thereby providing a transparent trajectory-level quality check rather than an additional ranking criterion.

\begin{figure}[t]
    \centering
    \includegraphics[width=\columnwidth]{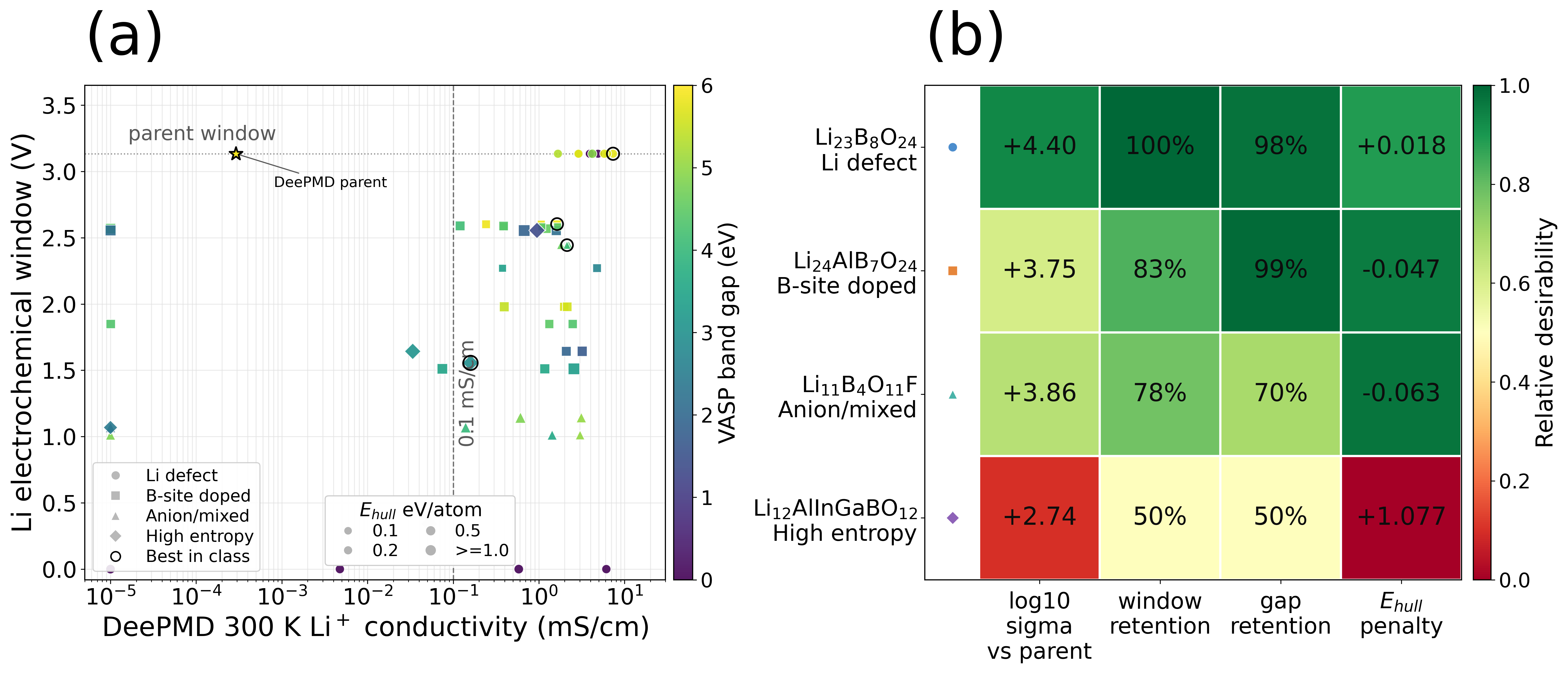}
    \caption{Trade-Offs Between Transport, Window, Band Gap, and Parent-Referenced Ehull for Li\textsubscript{3}BO\textsubscript{3} Derivative Candidates. (a) Global map of conductivity-window-band gap-parent-referenced Ehull for derivative structures. The horizontal axis represents 300 K Li\textsuperscript{+} conductivity, the vertical axis represents Li electrochemical window, point shapes represent Li defect, B-site doping, anion/mixed regulation, and high-entropy strategies, color represents band gap, point size represents parent-referenced Ehull, black circles represent the highest-scoring structure in each category, and the star represents the Li\textsubscript{3}BO\textsubscript{3} parent structure. (b) Comparison of metrics for the best integrated modification structure in each category relative to the Li\textsubscript{3}BO\textsubscript{3} parent structure. The numbers in the cells sequentially represent $\log_{10}\left(\sigma/\sigma_{\mathrm{parent}}\right)$, window retention rate, band-gap retention rate, and $\Delta E_{\mathrm{hull}}$. Color indicates relative desirability, with green being more favorable and red indicating higher costs.}
    \label{fig:figure8}
\end{figure}

Next, using the classic low-transport oxide Li\textsubscript{3}BO\textsubscript{3} as a benchmark~\cite{56}, the feasibility of local chemical modification for synergistically optimizing its conductivity and key properties was evaluated. We establish a chemically constrained, non-exhaustive structural design set consisting of 58 Li\textsubscript{3}BO\textsubscript{3} derivatives via hypothesis-driven perturbation strategies, including Li defect engineering, B-site doping, anion/mixed-anion regulation, and high-entropy configuration. Room-temperature conductivity, band gap, energy above hull, and electrochemical stability window were calculated using a unified standard. Fig.~\ref{fig:figure8}\textcolor{blue}{(a)} show that 46 out of the 58 modified structures have room-temperature conductivities exceeding 0.1 mS/cm, and 35 exceed 1.0 mS/cm, confirming that intrinsically low-transport oxides such as Li\textsubscript{3}BO\textsubscript{3} can be significantly activated by local chemical perturbations, with the best cases reaching more than four orders of magnitude improvement. However, different modification strategies exhibit markedly different performance gains and inherent trade-offs. Li defect engineering most effectively retains the parent phase’s wide electrochemical stability window and large band gap. It yields a maximum room-temperature conductivity of 7.34 mS/cm, but some high-defect-concentration configurations exhibit abrupt collapses in band gap or complete loss of electrochemical stability, accompanied by a slight elevation in the overall relative convex hull energy. B-site doping and anion/mixed regulation strategies enable conductivities exceeding 1.0 mS/cm in multiple structures. Notably, certain configurations simultaneously lower the relative convex hull energy and enhance thermodynamic stability, though they generally exhibit narrower electrochemical stability windows and reduced band gaps relative to the parent phase. High-entropy configurations can yield appreciable ionic conductivities but incur substantial performance penalties, including significant band gap narrowing, compressed electrochemical stability windows, and a pronounced deterioration in thermodynamic stability. Fig.~\ref{fig:figure8}\textcolor{blue}{(b)} compares representative Li\textsubscript{3}BO\textsubscript{3}-derived structures from four modification categories to illustrate the trade-off among Li-ion transport, electrochemical-window retention, band-gap retention, and thermodynamic stability. For the Li defect strategy, Li\textsubscript{23}B\textsubscript{8}O\textsubscript{24} exhibits the best balanced performance, delivering a 4.40-order-of-magnitude improvement in conductivity, nearly full retention of the electrochemical stability window and band gap, and a marginal increase of only 0.018 eV/atom in relative convex hull energy. The B-site doping representative Li\textsubscript{24}AlB\textsubscript{7}O\textsubscript{24} achieves a log\textsubscript{10} conductivity gain of +3.75, preserves 83\% of the electrochemical stability window and 99\% of the band gap, and reduces the convex hull energy by 0.047 eV/atom, thus enhancing its thermodynamic stability. For the anion/mixed regulation strategy, Li\textsubscript{11}B\textsubscript{4}O\textsubscript{11}F yields a log\textsubscript{10} conductivity improvement of +3.86, retains 78\% of the electrochemical stability window and 70\% of the band gap, and lowers the relative convex hull energy by 0.063 eV/atom, leading to a notable enhancement in thermodynamic stability. The high-entropy candidate Li\textsubscript{12}AlInGaBO\textsubscript{12} delivers the conductivity gain of +2.74 orders of magnitude, but retains only 50\% of both the electrochemical stability window and band gap, and its relative convex hull energy rises significantly, which translates to a prohibitively high overall performance cost.
\section{Discussion}\label{sec3}

In this work, three longstanding challenges in multi-constraint SSE screening—training data distribution mismatch, cross-property dataset heterogeneity, and kinetic data scarcity—are addressed via a hierarchical synergistic deep learning framework. This cascaded architecture deploys specialized models within their validated performance domains to balance computational efficiency, prediction accuracy, and screening reliability for ultra-large chemical spaces. In MP benchmark tests, the framework achieves significantly higher decision agreement than the same-protocol M3GNet workflow at critical screening nodes. Upon deployment to the 30.36-million Alex/ICSD-derived candidate space, 97 high-performance candidates are identified with room-temperature conductivities of 0.109 to 59.0 mS/cm.

Despite these significant advances, several limitations of the present work warrant acknowledgment. First, our structure generation is primarily based on elemental substitution of existing crystal prototypes, which inherently limits the discovery of entirely novel structural motifs. Second, some calculations were performed on single crystals at 0 K, ignoring critical factors under real battery operating conditions such as finite temperature effects, electrode-electrolyte interface reactions, grain boundaries, phase impurities, and microstructural defects~\cite{42,43,57}, leading to a notable discrepancy between computational predictions and experimental performance. Furthermore, the extrapolation capability of MLPs remains constrained, and prediction reliability deteriorates for materials whose chemical compositions lie outside the training distribution1~\cite{14,16,24}. Finally, the present gate-and-route sequence is specified a priori using physical reasoning and systematic model benchmarks. This controlled design improves transparency and limits error propagation, but it does not yet adapt the route automatically when the target property, underlying physics, chemical domain, data quality, model uncertainty, or computational budget changes. 

These limitations define five priorities for future development. First, designing novel structures with innovative anion frameworks and 3D continuous migration pathways to expand material exploration boundaries. Second, developing multi-scale coupled methods integrating interface chemistry, grain boundary transport and wide-temperature kinetics into a unified screening system to bridge the calculation-experiment gap. Third, optimizing MLPs via active learning and multi-fidelity fusion, and expanding cross-system training datasets to enhance model extrapolation in unexplored chemical spaces.  Furthermore, the current manually specified hierarchy should evolve into a property-aware mixture-of-experts architecture with learnable gating and routing. The gating model would evaluate the target property, candidate representation, chemical domain, label fidelity, predictive uncertainty, and computational budget, and then select the expert or sequence of experts best matched to each candidate and decision. Active learning, multi-fidelity fusion, and cross-system data expansion could update both the experts and the routing policy as new evidence becomes available. Such adaptive routing would support integrated high-throughput screening across thermodynamic, electronic, mechanical, electrochemical, and kinetic objectives while maintaining an explicit balance among throughput, accuracy, and reliability. Finally, extending the hierarchical screening framework to Na\textsubscript{+}/Mg\textsuperscript{2+} conductors and electrode interface protective layers for technical reuse and scalability.
\section{Data and Methods}\label{sec4}
\subsection{Candidate Structure Generation}
Candidate generation started from 81 elements. Noble gases were excluded owing to their chemical inertness; actinides and promethium were excluded owing to radioactivity; the late lanthanides (Sm–Lu) and several geologically scarce elements were excluded on cost and availability grounds; and transition metals with readily accessible redox couples were excluded because they tend to introduce electronic conductivity or parasitic redox activity in the bulk electrolyte. Electrochemically inert, high-valence cations that are established components of halide solid electrolytes, such as Sc\textsuperscript{3+}, Y\textsuperscript{3+}, and the early lanthanides La\textsuperscript{3+}–Nd\textsuperscript{3+}, were retained. To obtain structurally representative parent prototypes, the DIRECT sampling algorithm~\cite{58} was first applied to the DCGAT-3 dataset of the Alex computational materials database~\cite{12} to extract ordered, integer-occupancy structures suitable for DFT calculations. The 2022 release of the ICSD~\cite{7} was separately used as an experimental crystallographic prototype source. Structural prototypes were first extracted using the Pymatgen StructureMatcher tool~\cite{46}, followed by the application of elemental equivalent substitution rules to generate derivative structures. This workflow yielded an initial candidate set of 30,364,908 structures spanning the vast majority of crystallographic space groups and encompassing all seven crystal systems. The elemental composition and crystal structure distribution of the initial candidate library are presented in Fig. S7.
\subsection{Model Training and Validation}
Training data for the L-G-DCNN model included DFT-calculated formation energies from the Open Quantum Materials Database (OQMD) and the Materials Project formation-energy task distributed through Matbench~\cite{28,59}, covering an energy range of -15 to 10.0 eV·atom\textsuperscript{-1}. The training set intentionally included unrelaxed structures and metastable phases to prevent the model from exclusively learning the lowest-energy states of each composition, thereby minimizing false negatives for metastable candidates during the early coarse-screening stage. The multi-fidelity model built on DenseGNN~\cite{18} employed a hierarchical test-set design, with separate general, Li-containing, and SSE-specific subsets for thermodynamic-stability and band-gap evaluation while elastic-modulus labels were drawn from the computed elastic-property dataset~\cite{60}. Strict measures were implemented to prevent data leakage. All machine learning potentials used in benchmarking and high-throughput transport pre-evaluation adopted publicly available pre-trained weights without task-specific fine-tuning, ensuring a consistent assessment of their out-of-the-box transferability.
\subsection{Electrochemical Stability Window Calculation}
Electrochemical stability windows of candidate materials were calculated using the MatterSim framework~\cite{26}. First, candidate structures were loaded, parsed into Pymatgen Structure objects, and their chemical systems extracted. Next, energy entries and corresponding corrections for each chemical system were retrieved from the MP database~\cite{5}. Subsequently, structure optimization parameters were defined via MatterSim’s RelaxCalc class, and relaxation was performed to obtain optimized structures and energies. Combined with energy corrections, 0 K phase diagrams were constructed via Pymatgen’s PhaseDiagram module~\cite{46}. Using these phase diagrams, electrochemical properties in Li-based systems were evaluated. Briefly, the reference energy of elemental Li was established, and elemental decomposition analysis was conducted via Pymatgen’s ElementProfile method to track chemical potential changes across redox steps along the lithiation pathway~\cite{42,43,46}. The reduction potential was defined as the voltage where the mole fraction of reactive Li approaches zero, while the oxidation potential corresponded to the first Li deintercalation event. The electrochemical stability window was taken as the difference between these two potentials. Notably, the reported window errors reflect computational consistency within this 0 K phase diagram and correction framework, and do not directly equate to finite-temperature stability at practical battery interfaces.
\subsection{Ionic Conductivity Calculation}
Li-ion conductivities and apparent transport activation energies were evaluated using MD simulations combined with MSD analysis and Arrhenius fitting. Statistically representative supercells were first constructed, and MD simulations were performed at the target temperatures to record atomic trajectories. After unwrapping the trajectories under periodic boundary conditions, the Li-ion self-diffusion coefficient was obtained from the linear region of the Li self-MSD according to

\[
D = \frac{1}{6}
\lim_{t \to \infty}
\frac{d\left\langle r^2(t)\right\rangle}{dt}
\]

The Nernst--Einstein conductivity was then estimated as

\[
\sigma_{\mathrm{NE}}
= \frac{n q_{\mathrm{Li}}^{2} D}{k_{\mathrm{B}} T}
\]

where n is the Li-ion number density, $q_{Li}$ is the Li-ion charge, $k_{B}$ is the Boltzmann constant, and T is the absolute temperature. This relation assumes uncorrelated Li-ion motion~\cite{61}. Therefore, Haven-ratio-related diagnostics and the consistency between self-MSD and collective charge-MSD were used for trajectory quality assessment, rather than for direct conductivity correction. When multi-temperature MD data were available, apparent activation energies for Li-ion transport were obtained from Arrhenius fitting of the temperature-dependent diffusion coefficients or conductivities. Finite-size effects were mitigated by using $\geq$ 1200-atom supercells for room-temperature runs. Statistical reliability was assessed using fixed-window, block, and trajectory-quality diagnostics. A formal size-convergence study and independent replicate trajectories were not available for every candidate. The impacts of lattice thermal expansion and cation disorder on migration pathways were also accounted for. The pre-trained MatterSim-5M potential was used for high-throughput transport pre-evaluation. At elevated temperatures (500–1100 K), $>$900-atom supercells were employed for 1 ns simulations. For 300 K and 400 K, supercells were expanded to $>$1200 atoms, with 1 ns long-duration runs capturing migration behavior under realistic operating conditions. DeePMD was used for transport calculations of final candidates, with system-specific deep potentials built via the DP-GEN iterative workflow~\cite{35,36,37} and evaluated against the DFT energy, force, and stress benchmarks in Fig. S10–S12. Initial training data were generated by perturbing DFT-optimized structures. Phase space was explored via multi-temperature (100–1200 K) MD, with training samples dynamically added based on model prediction uncertainties until convergence. Nernst-Einstein relation is applied for conductivity estimation from self-diffusion coefficients, with Haven factor diagnostics performed for quality assessment (not used for conductivity correction). Transport coefficients were obtained from multiple-time-origin self-MSD averaging. For visualization and full-trajectory endpoint diagnostics in Fig. S9, PBC-unwrapped single-origin Li-MSD curves are also reported. They were not used as independent conductivity estimates. Trajectory quality is systematically evaluated through R² goodness-of-fit, linear-slope stability, and plateau-behavior inspection, with self-MSD and charge-MSD consistency used as a diagnostic for correlated ion motion. Training labels were obtained from DFT calculations using the PBE exchange-correlation functional~\cite{62}.

For full-cohort quality control, the final 97 trajectories were additionally processed using a common single-origin, PBC-unwrapped Li-MSD analysis and a density-site transition analysis (Fig. S8 and S9). The latter maps occupied Li sites and observed transitions in fractional coordinates and reports jump counts, the fraction of jumping Li ions, the largest connected-component fraction, and spanning axes. These diagnostics were used to inspect trajectory morphology and network connectivity. The final pass/fail decision remained based on the unified conductivity/MSD-consensus protocol described above.

\subsection{DFT Calculation Details}
All DFT calculations were performed with the Vienna Ab initio Simulation Package (VASP)~\cite{63} using the projector augmented-wave (PAW) method, with the PBE exchange-correlation functional as default. Convergence criteria were as follows: a plane-wave cutoff energy of 520 eV for accurate electronic state description. A k-point sampling density of 0.26/\AA for band structure convergence. And an electronic self-consistent field (SCF) threshold of 10\textsuperscript{-6} eV. For structural relaxation, forces were converged to $\leq$0.01 eV/\AA, with a total energy change of $<$1 meV/atom per ionic step.

To evaluate the fidelity of the potential over the configurational regimes sampled during transport analysis, VASP-PBE single-point energy, force, and stress labels were generated for 916 structures and compared directly with the corresponding DeePMD predictions (Fig. S10–S12). The validation set contains 441 frames from 300 K trajectories, 379 small-strain structures generated from 300 K final frames, and 96 high-temperature trajectory frames spanning 500–1000 K. The resulting energy, force-component, and stress-component MAEs are 0.00941 eV/atom, 0.02291 eV/\AA, and 5.61 × 10\textsuperscript{-4} eV/\AA\textsuperscript{3}, respectively, with R\textsuperscript{2} values of 0.9996, 0.9879, and 0.9659. This benchmark tests both near-ambient and thermally or mechanically perturbed configurations and provides the DFT-referenced energy/force/stress support for the long-time transport calculations.

\section{Data Availability}\label{sec5}

The data supporting the findings of this study are available within the Article and its Supporting Information. The submission-version code and processed dataset—including candidate lists and crystal structures, screening summaries, processed transport and DFT energy–force–stress validation data, BVSE cross-validation tables, and analysis and reproducibility scripts—are publicly available in the MatCascade repository at \textcolor{blue}{https://github.com/dhw059/MatCascade}. Owing to their size, the raw long-timescale molecular-dynamics trajectories, DFT input and output files, trained potential files, and the complete intermediate library of 30,364,908 expanded candidates are not hosted in the repository and may be obtained from the corresponding authors upon reasonable request, subject to applicable storage and licensing constraints. Source records from the Materials Project, the Open Quantum Materials Database, the Inorganic Crystal Structure Database, and the Alexandria database should be obtained from the respective providers in accordance with their access and licensing conditions.

\bibliography{sn-bibliography}

\end{document}